\documentclass{revtex4}
\usepackage[dvips]{color}
\usepackage{graphicx}
\usepackage{epsfig}

\begin{document}

\title{Non-adiabatic Chaplygin gas}
\author{H. A. Borges$^{a}$\footnote{humberto@ufba.br}, S. Carneiro$^a$\footnote{saulo.carneiro@pq.cnpq.br},
J. C. Fabris$^{b}$\footnote{fabris@pq.cnpq.br} and W. Zimdahl$^{b}$\footnote{winfried.zimdahl@pq.cnpq.br}}

\affiliation{$^{a}$ Instituto de F\'{\i}sica, Universidade Federal da Bahia, Salvador,
Bahia, Brazil\\ $^b$ Departamento de F\'{\i}sica, Universidade Federal do
Esp\'{\i}rito Santo, Vit\'oria,
Esp\'{\i}rito Santo, Brazil}


\begin{abstract}
The split of a generalised Chaplygin gas with an equation of state  $p = -A/\rho^{\alpha}$ into an interacting mixture of pressureless matter and a dark-energy component with equation of state $p_{\Lambda} = - \rho_{\Lambda}$ implies the existence of non-adiabatic pressure perturbations.
We demonstrate that the square of the effective (non-adiabatic) sound speed $c_s$ of the medium
is proportional to the ratio of the perturbations of the dark energy to those
of the dark matter. Since, as demonstrated explicitly for the particular case $\alpha = -1/2$, dark-energy perturbations are negligible compared with dark-matter perturbations
on scales that are relevant for structure formation, we find $|c_s^{2}| \ll 1$. Consequently, there are no oscillations or instabilities which have plagued previous adiabatic Chaplygin-gas models.
\end{abstract}

\maketitle

\section{Introduction}

The fact that apparently about $95\%$ of the matter-energy content of the Universe manifest themselves only through their gravitational effects \cite{wmap}, with no direct interaction with light, has led to many models trying to explain the physical nature of the cosmic substratum. This non-baryonic sector is commonly divided into two parts: dark matter (DM) with zero effective pressure, which is mainly concentrated in local structures
of the universe, and dark energy (DE) with negative pressure that drives the accelerated expansion and influences the anisotropy spectrum of the cosmic microwave background
radiation (CMB). The standard cosmological model incorporating these components is the $\Lambda$CDM model, which includes a zero-pressure fluid to represent DM and a cosmological constant $\Lambda$,
whose origin can be theoretically connected to the quantum vacuum energy. In general, this model is very successful but faces, at the same time, some important difficulties, either from
observational (excess of small structures, for example) or theoretical (the difficulty in deriving the quantum vacuum density) points of view.
\par
Many alternatives to the $\Lambda$CDM model have been proposed. Among them there are the interacting models, where the dark components interact directly (see, e.g., \cite{maartens,chimento,abdalla,annap,pavon} and references therein). These interacting models
intend to alleviate the so-called coincidence problem, related to the fact that both dark components have similar densities today, even if they scale very
differently with the expansion of the universe. Examples of interacting models are obtained by admitting a dynamical behavior of DE. In particular, decaying vacuum models have received interest since these models are in some cases competitive with the $\Lambda$CDM model \cite{saulo,cassio}.
\par
Another alternative is to incorporate both DE and DM into a single fluid, leading to unified models for
the dark sector.
The paradigm of unified models is the Chaplygin-gas model together with its generalisations \cite{uni,duni,te,salame}. The Chaplygin gas, defined by the equation of state (EoS)
$p = - A/\rho$, where $A$ is a constant and $\rho$ is the energy density, has a negative pressure and a positive sound speed. One of its extensions is
the generalised Chaplygin gas with an EoS $p = - A/\rho^\alpha$, where $\alpha$ is a free parameter. In some cases it is possible to map interacting models into the unified
framework. For example, a model in which a vacuum term decays linearly with the Hubble rate  \cite{saulo, cassio} corresponds, at the background level, to $\alpha = - 1/2$.
\par
Unified models of the dark sector are successful in describing the homogeneous and isotropic background dynamics of the Universe \cite{colistete,fabris}. However, at the perturbative level some
dangerous problems appear. In reference \cite{ioav} it has been remarked that the matter power spectrum for the generalised Chaplygin gas exhibits strong oscillations or instabilities, unless the model is in
a configuration that essentially reduces it to the $\Lambda$CDM model. Even if the observed matter power spectrum refers to the baryonic component, which may be well behaved, such oscillations
in the spectrum of the dark sector have undesirable consequences for the CMB spectrum \cite{finelli}. In reference \cite{rrrr} a solution to this problem has been proposed, making use of
entropic perturbations. Entropic perturbations modify the effective sound velocity of the fluctuations, damping or even suppressing completely the oscillations and instabilities in the spectrum of the dark
sector. However, no effective mechanism was proposed to justify the presence of non-adiabatic perturbations.
\par
On the other hand, progress was made on the basis of bulk-viscous models of the cosmic substratum.
Bulk-viscous models share the same homogeneous and isotropic background dynamics as generalised Chaplygin gases but, different from the latter, are characterised by non-adiabatic pressure perturbation.
In a sense, these models, which avoid the mentioned shortcomings, can be seen as ``non-adiabatic Chaplygin-gas models" \cite{VDF}.

Another possible mechanism to suppress the oscillations relies on a split of the Chaplygin fluid into two interacting components, one representing
DM and the other one the cosmological term \cite{bento}.
An alternative split  to generate entropic perturbations
was proposed in reference \cite{winfriedbis}.
In a recent paper \cite{wands}, entropic perturbations in a two-component formalism have been explored again to solve the oscillation problem. There, the absence of pressure perturbations was postulated by assuming that dark matter follows geodesics, equivalent to restrict the interaction to an energy transfer in the matter rest frame.
\par
In this work we come back to the issue of entropic perturbations in the generalised Chaplygin gas. We perform a decomposition of this gas by separating pressureless DM
from a cosmological term, both necessarily interacting with each other not only in the background but on the perturbative level as well. 
Cosmological perturbations in the resulting two-component system are intrinsically non-adiabatic.
The appearance of relevant entropic perturbations is a late-time effect, the perturbations being almost entirely adiabatic at early times. This fact allows us to fix the initial conditions
for the perturbations according to the standard model.
Our aim is to provide a physical reason for the expectation that non-adiabatic pressure perturbations may largely compensate the adiabatic contribution which is responsible for oscillations and/or instabilities.
On this basis the effective sound speed may then become negligible.
The crucial point is that our model implies a relationship between the effective sound speed
and the ratio of the energy-density perturbations of DE and DM.
In particular, we shall show that the smallness of the effective sound velocity is directly related to the smallness of the  DE perturbations compared with the matter perturbations.
For a model equivalent to the generalised Chaplygin gas with $\alpha = - 1/2$, it can be shown explicitly
that DE perturbations are indeed negligible on scales that are relevant for structure formation \cite{zimdahl}. As a consequence, the scale dependence in the perturbation equations becomes negligible
and neither oscillation nor instabilities occur.

\par
This paper is organized as follows. In Section \ref{background} we decompose the background dynamics of the generalised Chaplygin gas into an interacting mixture of pressureless matter and a time-varying cosmological term. In Section \ref{perturbations} a gauge invariant perturbative analysis
is performed and a relation between DE perturbations and the effective sound speed is derived. In Sections \ref{powerspectrum} and \ref{tests} observational parameter estimations are obtained at background and perturbative levels. In Section \ref{final} we present our conclusions.

\section{Background dynamics}
\label{background}

The Friedmann equation in a spatially flat universe is
\begin{equation}
3 H^{2} \equiv 3\bigg(\frac{\dot a}{a}\bigg)^2 = \rho\ ,\label{Friedmann}\\
\end{equation}
where units were fixed by imposing $8\pi G= c=1$ and $a$ is the scale factor of the Robertson-Walker metric. The  dot means derivative with respect to the cosmological time.
Additionally, we have local energy conservation
\begin{equation}
\dot\rho + 3\frac{\dot a}{a}(\rho + p) = 0 \label{continuity}.
\end{equation}
The pressure is given by
\begin{eqnarray}
\label{ed}
p = - \frac{A}{\rho^\alpha},
\end{eqnarray}
where $A$ is a positive constant, and $\alpha$ is a free parameter.
Using $(\ref{ed})$ in the continuity equation (\ref{continuity}), one obtains the solution
\begin{equation}\label{edof}
\rho=\left[A+\frac{B}{a^{3(1+\alpha)}}\right]^{1/(1+\alpha)},
\end{equation}
with an integration constant $B$. The present value of the scale factor was put to $a_{0}=1$.
The constant $B$ can be eliminated by defining $\bar{A}=A/\rho_0^{\alpha + 1}$. Throughout, the sub-index $0$ denotes the present value of the corresponding quantity.
Hence, the gas energy density is given by
\begin{equation}\label{ed3}
\rho = \rho_0\left[\bar{A}+\frac{1-\bar{A}}{a^{3(1+\alpha)}}\right]^{1/(1+\alpha)}.
\end{equation}
Note that for $\alpha = 0$ the gas reduces to a mixture of a cosmological constant and pressureless matter, with $\rho/\rho_0 = \bar{A}+(1-\bar{A})/a^3$.

The EoS parameter and the adiabatic sound velocity are
\begin{equation}\label{ed2}
\omega=\frac{p}{\rho} = - \bar{A}\left(\frac{\rho_0}{\rho}\right)^{\alpha + 1} \qquad \mathrm{and} \qquad
c_{a}^2=\frac{\dot{p}}{\dot{\rho}}=-\alpha\omega\ ,
\end{equation}
respectively.
For $\alpha >-1$, the energy density (\ref{ed3}) scales as pressureless matter for early times ($a \ll 1$) and
it behaves as a cosmological constant for $a \rightarrow \infty$.
Notice that $c_{a}^2$ is negative for negative values of $\alpha$.
The problems of (generalised) Chaplygin-gas models can be traced back to the circumstance that $|c_{a}^2|$ becomes of the order of unity unless $\alpha$ is extremely small, in which case its dynamics reduces to that of the $\Lambda$CDM model. For finite values of $\alpha$ the perturbation dynamics suffers from (unobserved) oscillations and/or instabilities \cite{ioav}. This is a consequence of the fact that the EoS (\ref{ed}) of the cosmic medium is of the type of an adiabatic EoS $p=p(\rho)$.
It has been suggested that a non-adiabatic perturbation dynamics may cure this shortcoming \cite{rrrr}.
Apart from adding non-adiabatic pressure perturbations ad hoc with the aim to cancel the unwanted adiabatic contributions, bulk viscous models were shown to give rise to a non-adiabatic perturbation dynamics in a natural way \cite{VDF}.
Here we present another way to avoid oscillations and/or instabilities for values of $\alpha$ not necessarily
very close to zero.
To this purpose, let us split this single fluid into two components, one of them with zero pressure.
Hence, we have
\begin{eqnarray}
\label{decomp}
\rho = \rho_m + \rho_\Lambda, \quad p = p_m + p_\Lambda
\end{eqnarray}
with
\begin{eqnarray} \label{decomp2}
p_m = 0, \quad p_\Lambda = - \frac{A}{\rho^\alpha}.
\end{eqnarray}
Furthermore, for the DE component we will generally (not only in the background)
assume an EoS
\begin{eqnarray} \label{decomp3}
p_\Lambda = - \rho_\Lambda\,,
\end{eqnarray}
corresponding to an energy-momentum tensor
\begin{equation}
T^{\mu}_{\nu} = \rho_{\Lambda} g^{\mu}_{\nu}.
\end{equation}
With the decompositions (\ref{decomp}), (\ref{decomp2}) and (\ref{decomp3}) the conservation equation (\ref{continuity}) takes the form
\begin{equation}\label{continuity2}
\dot\rho_m + 3\frac{\dot a}{a}\rho_m = - \dot\rho_\Lambda,
\end{equation}
and the DE component behaves as
\begin{eqnarray}\label{rho_L}
\rho_\Lambda = \frac{A}{\rho^\alpha} = \rho_0\bar A\bigg[\bar A + (1 - \bar A)a^{-3(1 + \alpha)}\bigg]^{-\frac{\alpha}{1 + \alpha}}.
\end{eqnarray}
The quantity $\bar A$ is identified as the present value of the DE density parameter,
\begin{equation}
\Omega_{\Lambda0} = \frac{\rho_{\Lambda0}}{\rho_0} = \bar A .
\end{equation}
Due to the decomposition (\ref{decomp})-(\ref{decomp3}) the pressureless matter component is obtained via
\begin{equation} \label{materia}
\rho_m = \rho - \frac{A}{\rho^{\alpha}}.
\end{equation}
With the help of equations (\ref{Friedmann}) and (\ref{rho_L}), it is easy to show that
\begin{equation}\label{ansatz}
\rho_{\Lambda} = \rho_{\Lambda 0} \left( \frac{H}{H_0} \right)^{-2\alpha}\ .
\end{equation}
The conservation equation (\ref{continuity2}) can be written in the form
\begin{equation}\label{continuity3}
\dot\rho_m + 3H\rho_m = \Gamma \rho_m,
\end{equation}
where
\begin{equation}\label{Gamma}
\Gamma = - 3\alpha \Omega_{\Lambda 0}H_{0} \left(\frac{H_{0}}{H}\right)^{\left(1+2\alpha\right)}
\end{equation}
is the rate by which the matter energy changes as a result of the interaction.
For $\alpha < 0$ the DE density decays along the expansion while DM is created.
In the particularly interesting case $\alpha = -1/2$ the DE density decays linearly with $H$ and matter is produced at a constant rate $\Gamma \sim H_0$ \cite{humberto}. On the other hand, for $\alpha = 0$ we re-obtain the standard model with a cosmological constant and conserved matter.

\section{Gauge-invariant perturbations}
\label{perturbations}

Using a spatially-flat background, the perturbed metric reads
\begin{equation}\label{fo}
ds^2= a(\eta)^2 [-(1+2\phi)d\eta^2+2B_{,i}d\eta dx^{i}+(1-2\psi)\delta_{ij}dx^{i}dx^{j}+2E_{,ij}dx^{i}dx^{j}],
\end{equation}
where just the scalar perturbations were retained.
In terms of the quantities $\phi$, $\psi$, $B$ and $E$ one defines the gauge-invariant Bardeen potentials
\begin{equation}\label{ha}
\Phi_{B}=\phi+\mathcal{H}(B-E')+(B-E')',
\end{equation}
\begin{equation}\label{ys}
\Psi_{B}=\psi-\mathcal{H}(B-E')\,,
\end{equation}
where the prime means derivative with respect to the conformal time $\eta$ and $\mathcal{H}=a'/a$.
Defining a velocity potential $v$ by $\delta u^{i} = a^{-1}\partial ^{i}v$ ($u^{\mu} = dx^{\mu}/ds$), where
$\delta u^{i}$ denotes the perturbed 4-velocity, suitable gauge-invariant perturbations of the matter quantities are
\begin{equation} \label{comoving}
\delta \rho^c = \delta \rho + \rho'\left(B + v\right), \quad \delta p^c = \delta p + p'\left(B + v\right),
\end{equation}
which are the pressure and density perturbations in the comoving frame.
Neglecting anisotropic stresses is equivalent to $\Phi_{B}=\Psi_{B}$.
Under this condition the  Bardeen potential satisfies the equation
\begin{equation}\label{tam}
\Phi_{B}''+3\mathcal{H}(1+c_{a}^2)\Phi_{B}'+[2\mathcal{H}'+(1+3c_{a}^2)\mathcal{H}^2+c_{s}^2k^2]\Phi_B = 0\,,
\end{equation}
where $k$ is the comoving wave-number and $c_s$ is the sound velocity of the cosmic medium as a whole, defined by
\begin{equation} \label{sound}
\delta p^c = c_s^2 \delta \rho^c\,.
\end{equation}
The Bardeen potential and the comoving density perturbation are related by the Poisson equation
\begin{equation}\label{op}
k^2\Phi_B=-\frac{a^2}{2}\delta\rho^{c}\,.
\end{equation}
Introducing $\delta\rho^{c}=\rho_m\delta^{c}$, where $\rho_m$ is the matter density (\ref{materia}), the Poisson equation $(\ref{op})$ takes the form
\begin{equation}\label{gf}
-2k^2\Phi_B=a^2\rho_m\delta^c.
\end{equation}
Substituting (\ref{gf}) and its derivatives in (\ref{tam}), we obtain a second order differential equation for $\delta^c$. If there where only adiabatic perturbations we would have $c_{s}^2=c_{a}^2$ with $c_{a}^2$
given by (\ref{ed2}).


So far, the perturbation equations are valid for the total energy density and the total pressure, irrespective of the decomposition into two components. Now, by using the split given in (\ref{decomp})-(\ref{decomp3}), we have, from (\ref{sound}),
\begin{equation}\label{sound2}
\delta p^c = \delta p_{\Lambda}^c = - \delta \rho_{\Lambda}^c = c_s^2 \delta \rho^c.
\end{equation}
There is no intrinsic non-adiabatic perturbation in the DE component.
Using $\delta\rho^{c} = \delta\rho^{c}_{\Lambda} + \delta\rho^{c}_{m}$, relation (\ref{sound2}) is
equivalent to
\begin{equation}\label{c_s}
\delta p^c  = - \delta \rho_{\Lambda}^c = \frac{c_s^2}{1+c_s^2} \delta \rho_m^c.
\end{equation}
Equation (\ref{c_s}) relates the perturbations of DE to those of DM via the sound speed of the cosmic medium. For $|\delta \rho_{\Lambda}^c| \ll |\delta \rho_m^c|$ we have $|c_s^2| \ll 1$  and vice versa.
Negligible DE perturbations imply a very small sound speed.

We mention that the expression (\ref{c_s}) is consistent with the general relation \cite{zimdahl}
\begin{equation} \label{winfried}
\delta p^c - \delta p^c_{ad} = \frac{\rho_{\Lambda}' \rho_m'}{\rho'} \left( \frac{\delta \rho_m^c}{\rho_m'} - \frac{\delta \rho_{\Lambda}^c}{\rho_{\Lambda}'} \right),
\end{equation}
which is obtained by combining (\ref{ed2}) and (\ref{sound2}). The right-hand side of eq.~(\ref{winfried})
represents the non-adiabatic part of the pressure perturbations which is due to the two-component nature of the system. In general, it does not vanish even if  each of the components is adiabatic on its own. It is this part which is supposed to largely cancel the adiabatic contribution $\delta p^c_{ad}$ with the result (\ref{c_s}) with $|c_{s}^{2}| \ll 1$.
Now, in any dynamic DE model inhomogeneities of the DE naturally appear. From the outset it is not clear that
these perturbations are small compared with the matter perturbations. Simply neglecting them may lead to
wrong interpretations of the observations \cite{Park-Hwang}. Whether or not DE perturbations are negligible compared with DM perturbations has to be studied on a case-by-case basis.
But, in fact, the decomposition (\ref{decomp})-(\ref{decomp3}) was introduced with the intention to separate
the clustering matter from the overall energy content of the Universe.
Here we focus on the choice $\alpha = -1/2$ that was previously studied in \cite{zimdahl}.
Under the assumption that (\ref{ansatz}) is the background version of the general solution
$\rho_{\Lambda} =  \rho_{\Lambda 0} \Theta/\Theta_0$, where $\Theta = u^{\mu}_{;\mu}$ is the fluid expansion, this case allows us
to obtain the DE perturbations explicitly in terms of the DM perturbations and their first derivative with respect to the scale factor,
\begin{equation}
\frac{\delta \rho^c_{\Lambda}}{\rho_{\Lambda}} = - \frac{1}{3K} \left(a \frac{\partial \delta^c_m}{\partial a} + \tilde{B} \delta^c_m \right)\,.
\end{equation}
Here,
\begin{equation}
K = 1 + \frac{1}{3} \tilde{A} - \frac{1}{6} \tilde{A} \tilde{B} - \frac{1}{9} \tilde{A}^{2}\,\frac{k^{2}}{a^{2}H^{2}}
\ ,
\label{K1}
\end{equation}
with
\begin{equation}
\tilde{A} = \frac{1 - \Omega_{m0}}{\ \Omega_{m0}a^{-3/2}}
\ ,\qquad
\tilde{B} = \frac{1 - \Omega_{m0}}{1 - \Omega_{m0} + \Omega_{m0}a^{-3/2}}
\ ,
\label{s/H}
\end{equation}
and $\delta_m^c = \delta \rho_m^c/\rho_m$.
The crucial point is the scale dependence of the factor $K$. While $\tilde{A}$ and $\tilde{B}$ are of the order of unity around the present time, one has $k^{2}/(a^{2}H^{2}) \gg 1$ on sub-horizon scales and, consequently, $|K|\gg 1$. It follows that for the present model we have indeed $|\delta \rho_{\Lambda}^c| \ll |\delta \rho_m^c|$ and $|c_s^2| \ll 1$.
Under this condition the scale dependence in eq.~(\ref{tam}) is negligible  and there appear neither oscillations nor instabilities in the power spectrum. The initial spectrum given at the end of the radiation era is just amplified during the Chaplygin-gas phase, almost entirely maintaining its shape.

\section{The power spectrum}
\label{powerspectrum}

\begin{figure*}
\vspace{.2in}
\centerline{\psfig{figure=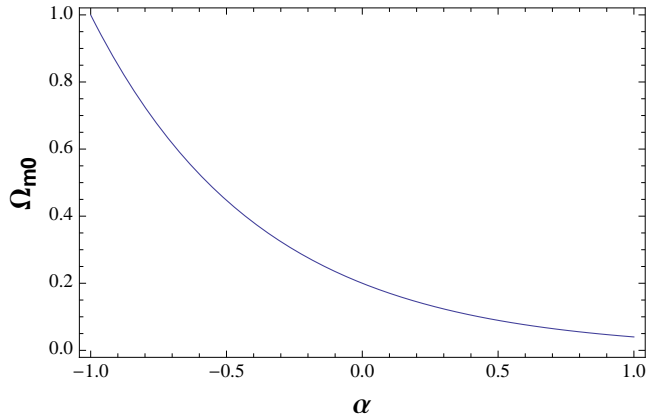}}
\caption{The present relative matter density $\Omega_{m0}$ as a function of the Chaplygin parameter $\alpha$.}
\label{fig1}
\end{figure*}
The power spectrum is characterised by the position of its turnover, which depends on $z_{eq}$, the redshift of matter-radiation equality. Let us show that the dependence of $z_{eq}$ on the present matter-density parameter $\Omega_{m0}$ depends on the Chaplygin-gas parameter $\alpha$ in a way that allows us to predict the best-fit value of $\Omega_{m0}$ for any $\alpha$ without having to construct the spectrum. As a particular example, we will show the spectrum for the case $\alpha = -1/2$, for which we also have a good concordance with the background observational tests.

Using equations (\ref{ed3}) and (\ref{materia}) and taking the limit $a \ll 1$, i.e. high redshifts, the matter density scales as
\begin{equation}\label{materia2}
\rho_m = 3 H_0^2 \Omega_{m0}^{\frac{1}{1+\alpha}} z^3 \qquad (z \gg 1).
\end{equation}
For $\alpha = 0$ we recover the standard result $\rho_m = 3 H_0^2 \Omega_{m0} z^3$. On the other hand, for $\alpha < 0$ we have, for the same present matter density, a lower amount of matter in the past. Equivalently, for the same amount of matter in the past, we will have more matter today. This results from the energy flux from DE to DM, encoded in (\ref{continuity3})-(\ref{Gamma}). Now, in order to locate the spectrum turnover at the correct position, we need the same amount of matter at $z_{eq}$ as in the standard model. Therefore, we expect a larger $\Omega_{m0}$ as compared to the standard case for $\alpha < 0$. Analogously, $\Omega_{m0}$ will be smaller than the standard value when $\alpha > 0$.

When only the power spectrum of large-scale structures (LSS) is taken into account, the best-fit value for the matter density parameter in the spatially-flat standard model ($\alpha = 0$) is given by $\Omega_{m0} \approx 0.2$, for both 2dFGRS and SDSS data \cite{2dF,SDSS}. By the way, this is in tension with the values $0.3 < \Omega_{m0} < 0.4$ obtained with supernovae analysis. As we shall see, this tension does not appear when we take $\alpha = -1/2$. From (\ref{materia2}) the redshift of radiation-matter equality is given by
\begin{equation}
z_{eq} =  \frac{\Omega_{m0}^{\frac{1}{1+\alpha}}}{\Omega_{R0}},
\end{equation}
where $\Omega_{R0}$ is the radiation density parameter today. Assuming the same amount of radiation as in the standard model, this redshift will also be the same, provided that
\begin{equation}
\Omega_{m0} \approx 0.2^{1+\alpha}.
\end{equation}
In Fig. \ref{fig1} we show this dependence of $\Omega_{m0}$ on $\alpha$. Models with $\alpha > 0$, i.e. with conversion of DM into DE, are clearly ruled out since they lead to matter densities below $0.2$, i.e., to a still higher tension with supernova observations. For $\alpha < 0$ we obtain larger values for the matter density, as discussed above.

\begin{figure*}
\vspace{.2in}
\centerline{\psfig{figure=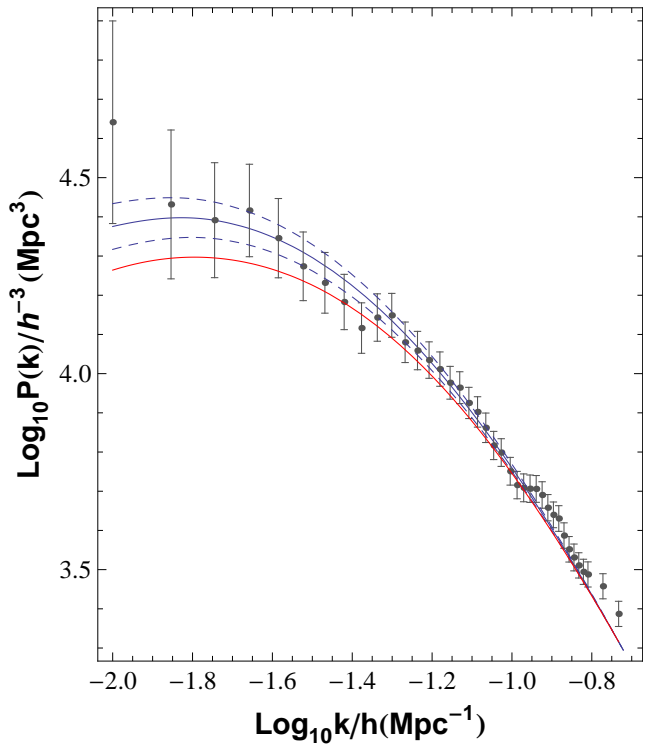}}
\caption{The power spectrum for $\alpha = -1/2$ and $\Omega_{m0} = 0.45$ (continuous blue line), and for the standard model best-fit (red line), with the 2dFGRS data points \cite{2dF}. The dashed lines delimit a $2\sigma$ confidence interval \cite{saulo}.}
\label{fig2}
\end{figure*}

In the particular case $\alpha = -1/2$ we have $\Omega_{m0} \approx 0.45$, and we will see that this is in agreement with the corresponding background tests. In Fig. \ref{fig2} we show the power spectrum obtained with this value. The continuous blue line corresponds to $\Omega_{m0} = 0.45$, the dashed lines delimit the $2\sigma$ confidence interval, and the red line represents the standard model best fit \cite{BBKS2}. This spectrum was originally obtained in reference \cite{julio}, by integrating the complete set of perturbed equations for matter and radiation from the deep radiation epoch to the present. We have imposed $\delta\rho_{\Lambda} \approx 0$, and used as initial conditions the Harrison-Zeldovich primordial spectrum. Since $\rho_{\Lambda} \ll \rho$ for early times, perturbations are almost adiabatic in this limit and we can use adiabatic initial conditions. In the integration we have assumed that baryons and dark matter follow the same trajectories, an approximation that, in the standard model case, leads to an error of about $10\%$. The spectrum was normalised by using small scale data, equivalent to use the observed value of $\sigma_8$. In other words, we have assumed that the ratio of the matter power spectrum to the galaxy power spectrum is equal to $1$. Other normalizations are possible, for example by using gravitational lensing data. This involves the study of quasi-linear and non-linear perturbations of the model, an issue currently under investigation \cite{chani}. Fig. 2 shows that at large scales the shape of the spectrum differs from the standard model prediction. This leaves a potential way to discriminate this model from $\Lambda$CDM.

\begin{figure*}
\vspace{.2in}
\centerline{\psfig{figure=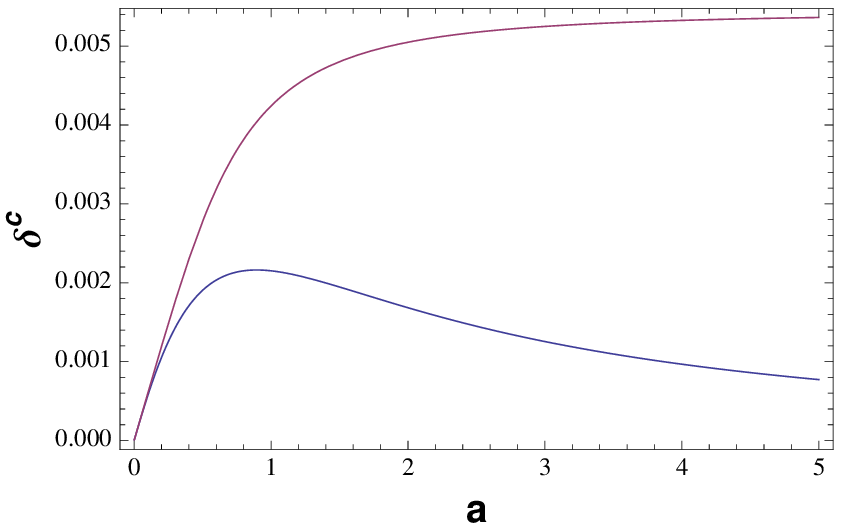} \hspace{.2in} \psfig{figure=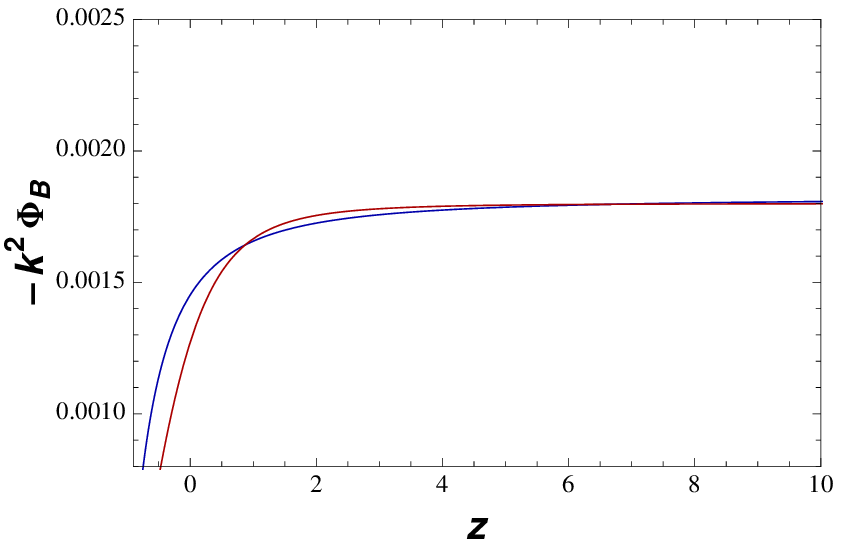}}
\caption{Left panel: The matter density contrast $\delta^c$ as a function of the scale factor for $\alpha = -1/2$ (blue line) and $\alpha = 0$ (red line) with the same initial conditions $\delta^c (z_{ls}) \approx 10^{-5}$ at the redshift $z_{ls}$ of last scattering. Right panel: The corresponding gravitational potentials as functions of the redshift.}
\label{fig3}
\end{figure*}

An interesting point to be shown is the evolution of the matter contrast along the universe expansion, given by the solution of (\ref{tam}) and (\ref{gf}) with negligible $c_s^{2}$. For $\alpha = -1/2$ and $\Omega_{m0} = 0.45$ it is depicted in the left panel of Fig. \ref{fig3} (blue line), together with the solution for $\alpha = 0$ and $\Omega_{m0} = 0.2$ (red line). We have used the same initial condition $\delta^c (z_{ls}) \approx 10^{-5}$ for both, where $z_{ls}$ is the redshift of last scattering. In the first case, the contrast approaches a maximum around the present time, i.e., when DM and DE have similar densities, and then decreases to zero as $a \rightarrow \infty$. The corresponding gravitational potential, given by (\ref{gf}), has essentially the same time evolution as in the standard model, as can be seen in the right panel of Fig. \ref{fig3}. For $z = 0$ the potentials differ only by about $10\%$. Any signature of this difference in the integrated Sachs-Wolf effect, for instance, would be masked by the cosmic variance.

\section{Joint observational tests}
\label{tests}

When performing background tests, we must take into account that matter is not conserved for $\alpha \neq 0$. For example, when testing the position of the first acoustic peak in the CMB anisotropy spectrum, the relation between the observed position $l_1$ and the acoustic scale $l_A$ is not the same as in the standard model. The general relation between these quantities is \cite{tegmark}
\begin{equation}
l_1 = l_A (1 - \delta_1),
\end{equation}
where
\begin{equation}
\delta_1 = 0.267 \left( \frac{r}{0.3} \right)^{0.1}, \qquad
r = \frac{\rho_R (z_{ls})}{\rho_m(z_{ls})}.
\end{equation}
In the last equation, the energy densities of radiation and matter are taken at the redshift of last scattering. With our result (\ref{materia2}) for the matter density at high $z$, we obtain
\begin{equation}
r = \Omega_{R0} z_{ls} \Omega_{m0}^{-\frac{1}{1+\alpha}}.
\end{equation}
For $\alpha = 0$ we have $r = \Omega_{R0} z_{ls} / \Omega_{m0}$, and only in this case the shift parameter of the flat standard model can be used.
On the other hand, when testing the distance to baryon acoustic oscillations, the parameter $\cal{A}$ introduced in \cite{eisenstein} can not be used either. It can only be used when the sound horizon radius scales with $\Omega_{m0}^{-1/2}$. In the general case it scales with $\Omega_{m0}^{-1/2(1+\alpha)}$, and we must explicitly use the scaling distance $D_V$ \cite{eisenstein} in the test.
Finally, when testing the Hubble diagram with type-Ia supernovas, some care is needed with the calibration procedure. One of the biggest current supernova compilations is the Union2 sample \cite{union2} the calibration of which is model-dependent since it uses a fiducial $\Lambda$CDM model with the Salt2 fitter. A Chaplygin model with fixed $\alpha$ has the same free parameters as the $\Lambda$CDM model, namely $H_0$ and $\Omega_{m0}$ and is not reducible to the latter. On the other hand, if we leave $\alpha$ free, the fitting of the Union2 sample will naturally lead to the $\Lambda$CDM best-fit with $\alpha \approx 0$. Therefore, the complementary use of model-independent compilations is mandatory, as for example the Constitution and SDSS samples, calibrated with the MLCS2k2 fitter \cite{MLCS}.



With these remarks in mind, we have performed a joint analysis of the case $\rho_{\Lambda} \propto H$ (that is, $\alpha = -1/2$), including the matter power spectrum, the position of the first peak in the CMB anisotropy spectrum, baryon acoustic oscillations and the three supernova samples referred to above \cite{saulo,cassio}. The results are summarised in Table I, together with the corresponding results for the $\Lambda$CDM model ($\alpha = 0$), after marginalisation of $H_0$. The values in each line correspond to the best-fit of the joint analysis of the corresponding SNe Ia compilation, CMB, BAO and the power spectrum. The resulting values of the reduced $\chi^2$ show that we have a good concordance for both models with the Union2 sample calibrated with Salt2. On the other hand, with the samples calibrated with the MLCS2k2 fitter we have a better concordance for $\alpha = -1/2$. The high values of the reduced $\chi^2$ obtained for $\Lambda$CDM indicate the tension present in that model between SNe Ia and LSS observations, as discussed above. For $\alpha = -1/2$, the resulting universe age for $H_0 \approx 70$ km/s$\cdot$Mpc is $t_0 \approx 13.5$ Gy.

\section{Final remarks}
\label{final}

As far as the homogeneous and isotropic background dynamics is concerned, the generalised Chaplygin gas serves as a prototype for unified models of the cosmological dark sector.
However, a successful description of structure formation requires a separation of the observable pressureless
matter component. A corresponding split of the total energy density of the cosmic substratum into DM and  DE is accompanied by an interaction between these components which in the case of the present paper amounts to a production of DM out of a decaying vacuum term.
Entropic perturbations do naturally appear in such a system and, different from previous approaches in the literature, do not have to be introduced ad hoc.
We have established a relation between the effective sound speed of the cosmic medium and the (scale-dependent) ratio
of the energy-density perturbations of DE and DM. According to this relation, the smallness of DE perturbations compared with matter perturbations implies the smallness of the effective sound speed.
For the special case $\alpha = - 1/2$ we obtain a value $|c_{s}^{2}| < 10^{-5}$
on typical scales \cite{zimdahl}. As a result, the scale-dependence of the perturbation equations is negligible.
There are no oscillations or instabilities in the matter distribution.
Different from the traditional adiabatic one-component description, the non-adiabatic Chaplygin gas may be a viable model of the cosmic substratum.
We obtain a good concordance when testing the model with $\alpha = -1/2$ against SNe Ia, BAO, CMB and LSS observations. This set currently constitutes the most precise and reliable observations we have. This concordance has also been confirmed by other complementary tests \cite{hermano}.
 This model has no $\Lambda$CDM limit.
 In the background the energy density ratio $\rho_{m}/\rho_{\Lambda}$ scales as $a^{-3/2}$, different from the $a^{-3}$ behavior of the corresponding quantity of the $\Lambda$CDM model, which amounts to an alleviation of the coincidence problem.
 In the late-time limit the matter density contrast is suppressed for the
Chaplygin gas, while it approaches a constant in the $\Lambda$CDM model.
Nevertheless, all the present results apply to the linear regime, and a study of the non-linear collapse is needed in the future. Such study is also necessary to verify the consequences of the contrast suppression on the formation of small-scale structures.
Another problem to be addressed is the full spectrum of CMB anisotropies. In the present work we have just considered the position of the first acoustic peak. Since we have the same matter density at high redshifts, the same content of baryons and radiation and the same time evolution of the gravitational potential as for the $\Lambda$CDM model, one may suggest that the same location of the first peak should be sufficient to show the equivalence of the spectra in both cases. But this has to be confirmed after adapting the current numerical codes to the case of two interacting and non-adiabatic fluids, which is not trivial. In particular, the creation of dark matter alters the relation between the present and past values of the matter density, a relation always implicit in the numerical codes. Another possible observation to be included in the parameter estimation in the future is weak lensing. It is expected to become a very powerful tool to discriminate dynamical models of dark energy \cite{dore}.

As a final comment we recall that in the particular case $\alpha = -1/2$ the DE density is linearly proportional to the Hubble parameter, $\rho_{\Lambda} = 2\Gamma H$.
Since such behavior is also expected for the QCD vacuum condensate in the expanding space-time with the correct order of magnitude for $\Gamma$ (see \cite{saulo} and references therein), this indicates a potential
microphysical foundation of our phenomenological model.
On this basis it would be natural to associate the produced DM particles with the condensate fluctuations.

\begin{table}[t]
\begin{center}
\caption{Limits to $\Omega_{m0}$ (SNe Ia + CMB + BAO + LSS), for $\alpha = -1/2$ \cite{saulo}.}
\begin{tabular}{rcccc}
\hline \hline \\
\multicolumn{1}{c}{ } & \multicolumn{2}{c}{Chaplygin gas} & \multicolumn{2}{c}{$\Lambda$CDM } \\
\multicolumn{1}{c}{SNe Ia sample}&
\multicolumn{1}{c}{$\Omega_{m0}$\footnote{Error bars stand for $2\sigma$.}}&
\multicolumn{1}{c}{$\chi^2_{min}/\nu$}&
\multicolumn{1}{c}{$\Omega_{m0}$$^a$}&
\multicolumn{1}{c}{$\chi^2_{min}/\nu$}\\ \hline \\
Union2 (SALT2)......&$0.420^{+0.009}_{-0.010}$ & 1.063 & $0.235\pm 0.011$ & 1.027 \\
SDSS (MLCS2k2).......& $0.450^{+0.014}_{-0.010}$ & 0.842 & $0.260^{+0.013}_{-0.016}$ & 1.231 \\
Constitution (MLCS2k2-17).......& $0.450^{+0.008}_{-0.014} $ &1.057 & $0.270\pm 0.013$ & 1.384\\
\hline \hline
\end{tabular}
\end{center}
\end{table}

\acknowledgments{The authors would like to thank Ioav Waga, Ribamar Reis and Hermano Velten for helpful discussions. This work was supported by CNPq (Brazil) and FAPES (Brazil).}

\end{document}